\documentclass[11pt]{article}
\usepackage{Moriond,epsfig}
\usepackage[usenames,dvipsnames]{color}
\usepackage{subfigure}
\RequirePackage[colorlinks=true
,urlcolor=blue
,anchorcolor=blue
,citecolor=blue
,filecolor=blue
,linkcolor=blue
,menucolor=blue
,pagecolor=blue
,linktocpage=true
,pdfproducer=medialab
]{hyperref}
\usepackage{cite}
\usepackage{amsmath}
\usepackage{mathrsfs}
\usepackage{amssymb} 
\usepackage{bbm}
\usepackage{slashed}

\newcommand{\tr}{\Tr}

\newcommand{\mean}[1]{\langle#1\rangle}

\newcommand{\ov}[1]{\overline{#1}}
\newcommand{\hc}{\text{h.c.}}
\newcommand{\unity}{\mathbbm{1}}

\newcommand{\LL}{\mathscr{L}}

\def\cA{{\cal A}}

\def\cY{{\cal Y}}
\def\cy{{\bf y}}
\def\Tr{{\rm Tr}}

\def\be{\begin{equation}}
\def\ee{\end{equation}}
\def\beq{\begin{equation}}
\def\eeq{\end{equation}}
\def\bc{\begin{center}}
\def\ec{\end{center}}
\def\bea{\begin{eqnarray}}
\def\eea{\end{eqnarray}}

\def\be{\begin{equation}}
\def\ee{\end{equation}}
\def\bea{\begin{eqnarray}}
\def\eea{\end{eqnarray}}

\begin{document}
\vspace*{4cm}
\title{NEUTRINO AND CHARGED LEPTON FLAVOUR TODAY~\footnote{Talk given by M.B.~Gavela at the XLVIIIth Rencontres 
de Moriond session devoted to ELECTROWEAK INTERACTIONS AND UNIFIED THEORIES, La Thuile (Italy), 2-9 March 2013.}}
\author{R.~Alonso$^1$, M.B.~Gavela$^{1,2}$, D.~Hern\'andez$^3$, L.~Merlo$^{1,2}$, and S.~Rigolin$^4$}

\address{$^1$Instituto de F\'{\i}sica Te\'orica UAM/CSIC and Departamento de F\'isica Te\'orica,\\
Universidad Aut\'onoma de Madrid, Cantoblanco, 28049 Madrid, Spain\\
$^2$CERN, Department of Physics, Theory Division CH-1211 Geneva 23, Switzerland\\
$^3$The Abdus Salam International Center for Theoretical Physics,\\ 
Strada Costiera 11, I-34151 Trieste, Italy\\
$^4$Dipartimento di Fisica ``G.~Galilei'', Universit\`a di Padova and \\
INFN, Sezione di Padova, Via Marzolo~8, I-35131 Padua, Italy}

\maketitle\abstracts{
Flavour physics is a priceless window on physics beyond the Standard Model. In particular, flavour violation 
in the lepton sector looks very promising, as high precision measurements are prospected in future experiments 
investigating on $\mu\rightarrow e$ conversion in atomic nuclei: the predictions for this observable are analysed 
in the context of the type I Seesaw mechanism. 
Furthermore, new ideas to explain the Flavour Puzzle recently appeared, mainly based on a possible dynamical 
origin of the Yukawa couplings and on flavour symmetries. The focus of this proceeding will be set on the 
Minimal Flavour Violation ansatz and on the role of the neutrino Majorana character: when an $O(2)_{N}$ 
flavour symmetry acts on the right-handed neutrino sector, the minimum of the scalar potential allows 
for large mixing angles -in contrast to the simplest quark case- and predicts a maximal Majorana phase. 
This leads to a strong correlation between neutrino mass hierarchy and mixing pattern.}

\section{Introduction}

Despite the Standard Model (SM) success~\cite{Aad:2012tfa,Chatrchyan:2012ufa}, experimental evidences for 
non-vanishing neutrinos masses, the presence of dark matter and the matter-antimatter asymmetry are calling 
for New Physics (NP) beyond the SM. Even within the theoretical context of the SM, puzzles such as the 
Hierarchy and the Flavour problems need as well an interpretation in terms of new particle physics.

At present, no evidence for non-standard particles beyond those in the SM spectrum has been claimed at colliders 
and this has a strong impact on theories beyond the SM, such as supersymmetry or models with a strongly interacting 
dynamics: the scale of new states/resonances expected in these contexts must be not lower than the TeV level 
(increasing in this way the intrinsic fine-tunings of those realizations). Nevertheless NP could also manifest 
indirectly through non-standard interactions, resulting in (still sizable) deviations from SM predictions for 
specific observables. This possibility has been deeply investigated in recent times both in the gauge-Higgs
~\cite{Giudice:2007fh,Grinstein:2007iv,Contino:2010mh,Alonso:2012jc,Azatov:2012bz,Corbett:2012dm,Espinosa:2012im,
Corbett:2012ja,Alonso:2012px,Alonso:2012pz,Azatov,Corbett:2013hia,Alonso:2013voa} and the flavour 
sectors~\cite{Ciuchini:2000de,Charles:2004jd}.

The presentation at the Moriond conference reviewed the prospects on Flavour Violation (FV) on the lepton sector: 
in particular, following Ref.~\cite{Alonso:2012ji}, the prospects of $\mu\to e$ conversion versus $\mu\to e \gamma$ and 
$\mu\to eee$ were presented and compared in the context of the type-I Seesaw mechanism for neutrino masses. 
Subsequently, the presentation discussed the theoretical attempts to understand the origin of the Flavour Puzzle 
in scenarios where the Yukawas have a dynamical origin. The main focus of this proceeding is however set on the latter. 

\section{Flavour violation in the lepton sector}

Future experiments aiming to detect  $\mu\rightarrow e$ conversion in atomic nuclei~\cite{Hungerford:2009zz,
Cui:2009zz,Carey:2008zz,Kutschke:2011ux,Kurup:2011zza} are especially promising for discovering FV in 
charged-lepton transitions and this motivates a dedicated investigation in the framework of the type-I 
Seesaw scenario (see~\cite{Alonso:2012ji} and references therein).

Analytically, the $\mu \to e$ conversion rate depends on form factors with and without a logarithmic dependence 
on the heavy singlet fermion masses. In Ref.~\cite{Alonso:2012ji}, this rate has been carefully computed and 
the results partially disagree with previous calculations: while there is an agreement on the logarithmic 
dependent terms (see for example Ref.~\cite{Deppisch:2010fr} and in Ref.~\cite{Ilakovac:2009jf} provided the 
non-supersymmetric limit of the quoted results is taken), there is however a disagreement on the other terms 
with calculations present in the literature. It is to be noticed that the constant terms in the form factors 
turn out to be numerically competitive with the logarithmic ones and therefore must be taken into account. 

\begin{figure}[h]
\centering
\includegraphics[width=0.4\textwidth]{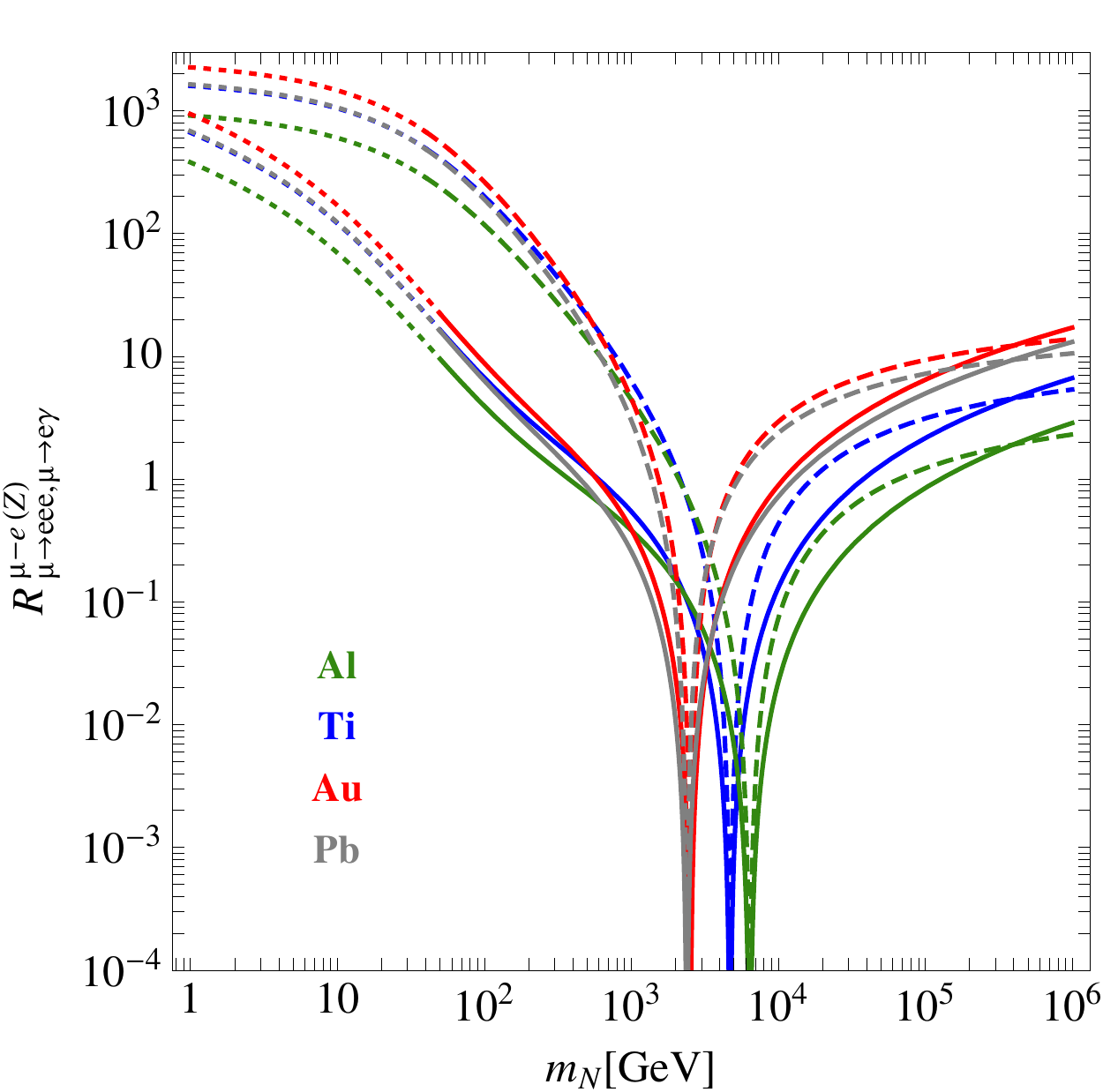}
\caption{Ratio of the $\mu\rightarrow e$ conversion rate in Al (green lines), Ti (blue lines) and Au (red lines) 
to  $Br(\mu\rightarrow e \gamma)$ (solid lines) and to $Br(\mu\rightarrow eee)$ (dashed lines) for the entire 
range in mass $m_N$ here considered. Lines are dotted when they require, for $\mu\rightarrow e \gamma$ and 
$\mu\rightarrow eee$, a sensitivity better than the one expected at planned experiments~\cite{Alonso:2012ji}.}
\label{figratiocomplete}
\end{figure}

The various possible ratios of rates involving the same charged $\mu-e$ flavour transition, have been determined 
in Ref.~\cite{Alonso:2012ji} and the results are summarized in Fig.~\ref{figratiocomplete}, presenting substantial 
differences with previous analysis in the literature. This is very useful for the comparison among the different 
processes within one same experiment and across experiments.

Interestingly, the ratios exhibit a different mass dependence and therefore can be useful to confirm/exclude 
scenario among the range of possible models. As an example, if the rates for $\mu\to e \gamma$ and $\mu\to e (Ti)$ 
are measured and their ratio is found to be $\sim 1$, then from Fig. \ref{figratiocomplete} the scale $m_N$ 
will be either $10^{3}$ GeV or $10^{4}$GeV, since there is a discrete degeneracy. Having determined the scale 
in this way, this type of models predicts the $\mu\to eee$ rate; in the given example, Fig.~\ref{figratiocomplete} 
signals a ratio among the rates of $\mu\to e (Ti)$ and $\mu\to eee$ of $\sim 10$ ($\sim 1$) for $10^{3}$ 
GeV ($10^{4}$ GeV), a prediction that could be checked against experiment. Note also that in the region $2-7$ TeV 
the conversion rate vanishes whereas $\mu\to e \gamma$ and $\mu\to eee$ are still observables.

The maximum scale for right-handed (RH) neutrino masses that future $\mu \to e$ conversion experiments could 
probe is above the $1000$ TeV scale. This sensitivity extends to very low masses, as low as $\sim2$ GeV for the 
Titanium case. As a result, planned  $\mu \to e$ conversion experiments will be fully relevant to detect or 
constrain scenarios with RH neutrinos in an impressive mass range.

\section{Dynamical Yukawas}

We move now to more theoretical aspects, reviewing some attempts done in the field for trying to shed some light on 
the origin of the Flavour Puzzle of the SM.

Yukawa couplings are the all and only source of flavour violation in the SM, where they arise at the renormalisable 
level. In beyond the SM frameworks one can follow two different approaches. From an effective field point of view, the 
Yukawa couplings can be though as arising from higher dimensional operators once heavier states are integrated out. 
The best known example is the Seesaw mechanism, in its different variants, where integrating out the heavier 
Majorana neutrino states resulted into the Weinberg effective operator at large energies. In this case, the Yukawa 
couplings explicitly break the flavour symmetry as in the SM.

Alternatively, one can take the picture in which Yukawa couplings correspond to dynamical fields of the fundamental 
theory. Once these fields develop vacuum expectation values (vevs) in the flavour space, a spontaneous breaking of 
the flavour symmetry occurs. The first attempt in this direction has been proposed in Ref.~\cite{Froggatt:1978nt} 
by Froggatt and Nielsen: a global $U(1)$ symmetry was assumed to act horizontally on the different fermion families; 
the SM spectrum is then enriched by a scalar field that only transforms under the flavour $U(1)$, usually called 
flavon; the Yukawa couplings arise as effective couplings written in terms of the vev of the flavon over the cutoff 
of the theory. In this way, the top quark mass appears at the renormalisable level, while all the other fermion masses 
originate from higher dimensional operators. In this light fermion masses suppression, relatively to the top scale, 
can be explained. The Froggatt-Nielsen model can indeed describe fermionic masses and mixings in agreement with  
present observations (see Refs.~\cite{Buchmuller:2011tm,Altarelli:2012ia} for a recent discussion), but suffers in 
general from the appearance of too large flavour changing neutral currents (FCNCs)~\cite{Calibbi:2012yj,Calibbi:2012at}.

The Froggatt-Nielsen proposal is at the origin of an abundant literature: the flavour symmetry $G_f$ has been 
considered gauged/global, continuous/discrete, Abelian/non-Abelian. In particular, models based on discrete 
symmetries received attention in the last years (see Refs.\cite{Altarelli:2010gt,Ishimori:2010au,Grimus:2011fk,
Altarelli:2012ss,Bazzocchi:2012st} for reviews), due to their ability in predicting specific neutrino mixing 
patters, such as the Tri-Bimaximal mixing. The Lagrangian of such models is invariant under a certain discrete 
symmetry and accounts for a scalar potential that is adjusted to spontaneously break $G_f$, leaving two different 
subgroups preserved, one in the neutrino sector and the other in the charged lepton one. It is the mismatch of these 
two subgroups that leads to the Tri-Bimaximal pattern. The main advantages of these models are their predicting 
power, the absence of Goldstone bosons due to the symmetry breaking and the suppression of FCNCs as discussed 
in Refs.~\cite{Feruglio:2007uu,Feruglio:2008ht,Ishimori:2008au,Ishimori:2009ew,Feruglio:2009iu,Feruglio:2009hu,Toorop:2010ex,Toorop:2010kt,Merlo:2011hw,Altarelli:2012bn}. 
Despite of their success, however, the downsides of the discrete symmetry approaches are that there is no rationale 
for choosing a particular discrete group, that the neutrino mixing and spectrum are not correlated, that it is 
difficult to account for leptons and quarks simultaneously, and finally that the relatively large measured value 
for the reactor angle $\theta_{13}$. Recent developments on discrete symmetry models can be found in 
Refs.~\cite{Altarelli:2009gn,Toorop:2010yh,Varzielas:2010mp,Ge:2011ih,Ma:2011yi,Meloni:2011fx,Morisi:2011pm,
Toorop:2011jn,King:2011zj,Hernandez:2012ra,Feruglio:2012cw,Holthausen:2012dk,Hernandez:2012sk,Feruglio:2013hia}.

Continuous flavour symmetries have also been deeply investigated, but mainly connected to the Minimal Flavour 
Violation (MFV)~\cite{Chivukula:1987py,Hall:1990ac} ansatz. In shorts, MFV is nothing else the requirement that 
all sources of flavour violation in the SM and beyond the SM is described at low-energies uniquely in terms 
of the known fermion masses and mixings. MFV emerged in the last years clearly as the most promising working 
framework consistent with the extremely stringent FCNC constraints~\cite{D'Ambrosio:2002ex,Cirigliano:2005ck,
Davidson:2006bd,Kagan:2009bn,Gavela:2009cd,Feldmann:2009dc,Alonso:2011yg,Alonso:2011jd,Alonso:2012fy}. Indeed, 
several distinct models based on the MFV ansatz~\cite{Lalak:2010bk,Fitzpatrick:2007sa,Grinstein:2010ve,
Buras:2011wi,Barbieri:2011ci,Alonso:2012jc,Blankenburg:2012nx,Alonso:2012pz,Lopez-Honorez:2013wla} are still consistent with a NP scale at the TeV, 
while comparable models without the MFV hypothesis are forced to have a scale larger than hundreds of 
TeV~\cite{Isidori:2010kg}. 

The power of MFV descends from the fact that it exploits the symmetries that the SM itself contains in a 
certain limit: that of massless fermions. For example, in the case of the Type I Seesaw mechanism with three 
RH neutrinos added to the SM spectrum, the flavour symmetry of the full Lagrangian, when Yukawa couplings and 
the RH neutrino masses are set to zero, is: 
\beq
G_f=G^q_f\times G^\ell_f\qquad\text{with}\qquad
\begin{cases}
G^q_f=U(3)_{Q_L}\times U(3)_{U_R}\times U(3)_{D_R}\\
G^\ell_f=U(3)_{\ell_L}\times U(3)_{E_R}\times U(3)_{N}\\ 
\end{cases}\,.
\label{GfGlobal}
\eeq
Under the flavour symmetry group $G_f$ fermion fields transform as
\beq
\begin{gathered}
Q_L\sim(3,1,1)_{G_f^q}\,,\qquad\qquad
U_R\sim(1,3,1)_{G_f^q}\,,\qquad\qquad
D_R\sim(1,1,3)_{G_f^q}\,,\\
\ell_L\sim(3,1,1)_{G_f^\ell}\,,\qquad\qquad
E_R\sim(1,3,1)_{G_f^\ell}\,,\qquad\qquad
N_R\sim(1,1,3)_{G_f^\ell}\,.
\end{gathered}
\label{FermionTransf}
\eeq
The Yukawa Lagrangian for the Type I Seesaw mechanism, then, reads:
\beq
-\LL_Y=\ov{Q}_LY_DHD_R+\ov{Q}_LY_U\tilde{H}U_R+\ov{\ell}_LY_EHE_R+\ov{\ell}_LY_\nu\tilde{H}N_R+
       \ov{N}^c_R\dfrac{M_N}{2}N_R+\hc
\label{Lagrangian}
\eeq 
To introduce $\LL_Y$ without explicitly breaking $G_f$, the Yukawa matrices $Y_i$ and the mass matrix for the 
RH neutrinos $M_N$ have to be promoted to be spurion fields transforming under the flavour symmetry as:
\beq
\begin{gathered}
Y_U\sim(3,\bar3,1)_{G_f^q}\,,\qquad\qquad
Y_D\sim(3,1,\bar3)_{G_f^q}\,, \hspace{4.5cm}\\
 Y_E\sim(3,\bar3,1)_{G_f^\ell}\,,\qquad\qquad
Y_\nu\sim(3,1,\bar3)_{G_f^\ell}\,,\qquad\qquad
M_N\sim (1,1,\bar6)_{G_f^\ell}\,.
\end{gathered}
\label{YTransf}
\eeq
The quark masses and mixings are correctly reproduced once the quark spurion Yukawas get background values as
\beq
Y_U =V^\dag\,\cy_U\,,\qquad\qquad
Y_D =\cy_D\,,
\eeq
where $\cy_{U,D}$ are diagonal matrices with Yukawa eigenvalues as diagonal entries, and $V$ a unitary matrix 
that in good approximation coincides with the CKM matrix. For lepton masses and mixings the discussion is more 
involved and it is postponed to Sect.~\ref{Sec:Leptons}. 

Despite of the phenomenological success, it has to be noticed that, however, MFV does not provide byh itself 
any explanation of the origin of fermion masses and/or mixing, or equivalently does not provide any explanation 
for the background values of the Yukawa spurions. This observation motivates the studies performed in 
Refs.~\cite{Alonso:2011yg,Alonso:2012fy} (see also Refs.~\cite{Anselm:1996jm,Barbieri:1999km,Berezhiani:2001mh,Harrison:2005dj} for earlier attempts towards a dynamical origin of the Yukawa couplings), where the Yukawa spurions are promoted to dynamical scalar fields: the case in which a one-to-one 
correlation among Yukawa couplings and fields is assumed, $Y_i\equiv \mean{\cY_i}/\Lambda_f$, is discussed at length. 
Moreover, other possible choices, such as for example $Y_i\equiv \mean{\chi^1_i}\mean{\chi^2_i}/\Lambda^2_f$, are 
also considered. The scalar potential constructed out of these fields was studied in Refs.~\cite{Alonso:2011yg,
Alonso:2012fy}, considering renormalisable operators (and adding also lower-order non-renormalisable 
terms for the quark case): these effective Lagrangian expansions are possible under the assumption that the ratio of the flavon vevs 
and the cutoff scale of the theory is smaller than 1, condition that is always satisfied but for the top Yukawa 
coupling. Int this case a non-linear description would be more suitable.

It turns out that the Majorana nature of neutrinos has a deep impact on the results: in Ref.~\cite{Alonso:2012fy}, 
it was analysed a particular Type I SeeSaw model with two degenerate RH neutrinos, corresponding to an $O(2)_{N}$ flavour symmetry for the right-handed neutrino sector; at the minimum of the scalar potential a large mixing 
angle -in contrast to the simplest quark case- and a maximal Majorana phase are predicted. This lead to a 
strong correlation between neutrino mass hierarchy and mixing pattern. In the following, we will show~\cite{Alonso:2013mca} that this result can be generalized to generic type I Seesaw models.

\subsection{The quark sector}

In Ref.~\cite{Alonso:2011yg}, it has been considered the case in which the quark Yukawa spurions correspond 
to dynamical fields of some fundamental theory, developing vevs through the minimization of a specific scalar 
potential (see also Refs.~\cite{Anselm:1996jm,Berezhiani:2001mh})\footnote{In order to avoid the appearance of 
Goldstone bosons, corresponding to the spontaneous breaking of the flavour symmetry, $G^q_f$ can be gauged. See 
Refs.~\cite{Grinstein:2010ve,Feldmann:2010yp,Guadagnoli:2011id,Buras:2011zb,Buras:2011wi}}. 

The renormalisable scalar potential for the quark Yukawa fields $\cY_U, \cY_D$ depends only on five\footnote{Here we 
are assuming that $G_f$ contains the $U(3)_i$ factors. Consequently the operators $\det\left(\cY_i\right)$ are not 
invariants, differently from what considered in Ref.~\cite{Alonso:2011yg}, where the flavour symmetry contained 
instead only the $SU(3)_i$ part. The effects of $\det\left(\cY_i\right)$ in the potential are to push towards 
degenerate mass configurations~\cite{Alonso:2011yg}.} independent invariant~\cite{Feldmann:2009dc,Alonso:2011yg}:

\beq
\begin{gathered}
\tr\left[\cY_U\cY_U^\dagger\right]\,,\qquad
\tr\left[\cY_D\cY_D^\dagger\right]\,, \qquad 
\tr\left[\left(\cY_U\cY_U^\dagger\right)^2\right]\,, \qquad 
\tr\left[\left(\cY_D\cY_D^\dagger\right)^2\right]\,, \\
\tr\left[\cY_U\cY_U^\dagger\cY_D\cY_U^\dagger\right]\,.
\end{gathered}
\label{QuarkInvariants}
\eeq

The terms in the first line of Eq.~(\ref{QuarkInvariants}) turn out to be responsible for fixing the quark mass 
hierarchies (see Ref.~\cite{Nardi:2011st,Espinosa:2012uu} for an alternative approach), while the term in the 
second line is the only involving the mixing angle $\theta$. Considering as exemplification the two-family case, 
(as the three family case is conceptually similar but with more complicate expressions), one obtains that  
\beq
\tr\left(\cY_U\cY_U^\dagger\cY_D\cY_D^\dagger\right)\propto \left(m_c^2-m_u^2\right)\left(m_s^2-m_d^2\right)\cos2\theta 
\,.
\eeq
Minimizing with respect to $\theta$, the following condition must holds:
\beq
\left(m_c^2-m_u^2\right)\left(m_s^2-m_d^2\right)\sin2\theta=0.
\eeq
There are only two possible solutions to the previous equation: either the same type quarks are degenerate or the 
mixing angle is vanishing. Only the second solution is a good (first order) approximation of the data. Then a Cabibbo 
like angle could be generate, for example, from sub-leading (higher order) terms. Considering non-renormalisable terms 
in the scalar potential, one notice that the first one containing information on the mixing angle can arise only at 
$d=8$. The scalar potential minimization predicts a mixing angle of the order of 
\beq
\sin^2\theta\simeq\dfrac{c}{2\,y_c^2\,y_s^2}
\eeq
with $c$ is a free (order one) parameter. A reasonable value for the Cabibbo angle is recovered only if $c \sim 
10^{-10}$, clearly not natural in an effective Lagrangian approach. 

In conclusion, the promotion of the quark Yukawas spurions to dynamical fields transforming in the bi-fundamental 
of the flavour group $G_f$ does not lead to a successful description of the quark mixing, for two or three families, not 
even if non-renormalisable terms are included. On the other side, if the Yukawas are thought as effective terms 
coming form the dynamics of scalar fields transforming in the fundamental of $G_f$, it is indeed possible to recover 
a hierarchical spectrum among two families and a mixing angle~\cite{Alonso:2011yg}. However, this strategy does 
not lead to a realistic description of the three-generation case, unless one introduces at the same time fields 
transforming in the fundamental and in the bi-fundamental, loosing however a direct connection between Yukawa couplings and flavons.

\subsection{The lepton sector}
\label{Sec:Leptons}

When discussing the MFV ansatz in the leptonic sector one has at disposal three different spurions, as can be 
evinced from the list in Eq.~(\ref{YTransf}). The number of parameters that can be introduced in the model through 
these spurions is much larger than the low energy observables. This in general prevents a direct link among neutrino 
parameters and FV observables. The usual way adopted in the literature to lower the number of parameters consists 
in reducing the number of spurions fron three to two: for example in Ref.~\cite{Cirigliano:2005ck} $M_N\propto\unity$ 
is taken; in Ref.~\cite{Gavela:2009cd}, a two-family RH neutrino model is considered with $M_N\propto\sigma_1$; 
finally in Ref.~\cite{Alonso:2011jd} $Y^\dag_\nu Y_\nu\propto\unity$ is assumed. 

An unifying description for all these models can be obtained by introducing the Casas-Ibarra parametrization 
\cite{Casas:2001sr}: in the basis of diagonal mass matrices for RH neutrinos, LH neutrinos and charged 
leptons, the neutrino Yukawa coupling can be written as 
\beq
Y_\nu=\dfrac{1}{v}U\sqrt{\hat m_\nu}R\sqrt{\hat M_N},
\eeq
where $v$ is the electroweak vev, the hatted matrices are light and heavy neutrino diagonal mass matrices, $U$ refers 
to the PMNS mixing matrix and $R$ is a complex orthogonal matrix, $R^T R=\unity$. A correct description of lepton 
masses and mixings is achieved assuming that $Y_E$ acquires a background value parametrised by a diagonal matrix,
\beq
Y_E=\cy_E\equiv{\rm diag}(y_e,\,y_\mu,\,y_\tau)\,,
\eeq
while the remaining spurion, $M_N$ or $Y_\nu$, accounts for the neutrino masses and the PMNS matrix (see 
Refs.~\cite{Cirigliano:2005ck,Gavela:2009cd,Alonso:2011jd}). 

In Ref.~\cite{Alonso:2012fy}, the simple model with 
only two heavy RH neutrinos is considered~\cite{Gavela:2009cd}. The spurion fields $Y_E$ and $Y_\nu$ are promoted to dynamical fields, 
$\cY_E$ and $\cY_\nu$, and the corresponding scalar potential is studied. As for the quark case, the scope of the  
analysis was to explain the origin of the spurion background values, necessary to correctly describe the measured lepton masses 
and mixings.

In the following, still focusing to a two-flavon model, we first generalise the analysis presented in 
Ref.~\cite{Alonso:2012fy} by considering a generic mass matrix for the RH neutrinos. Then we will discuss the 
limit of degenerate RH neutrino masses. Finally, we briefly explore also the case in which a third flavon, the 
RH neutrino mass, is introduced. Through all the discussion we will first concentrate in the two-family case and 
only subsequently we will generalise the results to three families. 


\boldmath
\subsubsection{Generic RH neutrino masses: $G^\ell_f=U(2)_{\ell_L}\times U(2)_{E_R}$.}
\unboldmath

A generic RH neutrino mass matrix breaks explicitly the $U(2)_{N}$ factor. As a result, the flavour symmetry of the kinetic terms reduces to
\beq
G_f^\ell=U(2)_{\ell_L}\times U(2)_{E_R}\,.
\label{GfGenericN}
\eeq
Fermion and flavon fields transform under $G_f^\ell$ as in the two-family version of Eqs.(\ref{FermionTransf}) 
and (\ref{YTransf}), but neglecting the transformations under the RH neutrino symmetry factor $U(2)_{N}$, that 
is explicitly broken. Alike to the quark case, only five independent invariants can 
be obtained at the renormalisable level:
\beq
\begin{aligned}
&\tr\left[\cY_E\cY_E^\dagger\right]\,,\qquad
&&\tr\left[\cY_\nu\cA \cY_\nu^\dagger\right]\,,\qquad 
&\tr\left[\left(\cY_E\cY_E^\dagger\right)^2\right]\,, \qquad 
&&\tr\left[\left(\cY_\nu\cA \cY_\nu^\dagger\right)^2\right]\,,\\
&\tr\left[\cY_E\cY_E^\dagger\cY_\nu\cA \cY_\nu^\dagger\right]\,.
\end{aligned}
\label{LeptonInvariants}
\eeq
The possibility of including a generic $2\times2$ matrix, $\cA$, in the neutrino invariants is a novelty compared 
to the quark case and it is possible thanks to the transformation properties of the neutrino Yukawa flavon, 
$\cY_\nu\sim(3,1)$. Analogously in the quark case, starting from the five invariants in Eq.~(\ref{LeptonInvariants}) it is 
build the corresponding renormalisable scalar potential. Terms in the first line account for the lepton masses 
while the operator in the second line fixes the mixing angle:
\beq
V_{\rm mix}=\tr\left[\cY_E\cY_E^\dagger\cY_\nu\cA\cY_\nu^\dagger\right]\propto \,
\tr\left[\cy^2_E\, U\,\sqrt{\hat m_\nu}\, P\, \sqrt{\hat m_\nu}\, U^\dag\right]\,,
\label{LeptMixInv}
\eeq
with $P\equiv R\, \sqrt{\hat M_N}\cA\sqrt{\hat M_N}\, R^\dag$ and $U$ being the two-family PMNS matrix defined by
\beq
U=\left(
\begin{array}{cc}
 \cos\theta & \sin\theta  \\
 -\sin\theta &  \cos\theta \\
\end{array}
\right)\left(
\begin{array}{cc}
 e^{-i\alpha} &   \\
  & e^{i\alpha}  \\
\end{array}
\right)\,.
\eeq 
Minimising the potential term in Eq.~(\ref{LeptMixInv}) with respect the angle $\theta$ and the Majorana phase $\alpha$, the following two conditions result:
\beq
\begin{aligned}
&2(y_\mu^2-y_e^2)\sqrt{m_1\,m_2}\sin2\theta\left|P_{12}\right|\sin(2\alpha-\arg P_{12})=0\,,\\
&(y_\mu^2-y_e^2)\Big[\sin2\theta\left(m_1\,P_{11}-m_2\,P_{22}\right)-2\cos2\theta\sqrt{m_1\,m_2}
\left|P_{12}\right|\cos\left(2\alpha-\arg P_{12}\right)\Big]=0\,.
\end{aligned}
\label{min2pot}
\eeq
Beside trivial configurations, Eq.~(\ref{min2pot}) allows the following solution~\cite{Alonso:2013mca}:
\beq
\begin{cases}
&2\alpha-\arg P_{12}=n\pi\qquad\text{with}\qquad n\in  \mathbb{Z}\\
&\tan2\theta=2\left|P_{12}\right|\dfrac{\sqrt{m_1\,m_2}}{m_1\,P_{11}-m_2\,P_{22}}\,.
\end{cases}
\label{LeptonSolutions}
\eeq
The first expression connects the low-energy and the high-energy phases, while the second one represents a link 
among the size of mixing angle and the type of the neutrino spectrum. It is the Majorana neutrino character that allows this novel connection. However, the presence of the generic matrix $\cA$ prevents the possibility of making clear predictions for the mixing angle.

\boldmath
\subsubsection{Degenerate RH neutrino masses: $G^\ell_f=U(2)_{\ell_L}\times U(2)_{E_R}\times O(2)_{N}$.}
\unboldmath

In the case with degenerate RH neutrino masses, $M_1=M_2\equiv M$, the flavour symmetry is
\beq
G_f^\ell=U(2)_{\ell_L}\times U(2)_{E_R}\times O(2)_{N}\,.
\eeq
This is the largest possible symmetry for the RH neutrino sector, once non-vanishing masses for RH neutrinos are considered.

The independent invariants~\cite{Alonso:2013mca}, at the renormalisable level, still include those in Eq.~(\ref{LeptonInvariants}), 
but since $\cY_\nu\sim(2,1,\bar 2)$, insertions of a generic matrix $\cA$ is not allowed anymore: the only 
possible choice for building operators invariants under $G_f^\ell$ are $\cA=\unity$ and $\cA=\sigma_2$. Among the 
three invariants that can be written with the insertion of $\cA=\sigma_2$, only one is not vanishing,
\beq
\tr\left[\left(\cY_\nu\sigma_2\cY_\nu^\dagger\right)^2\right]\,.
\eeq
An equivalent way to rephrase this operator, without the explicit presence of $\sigma_2$, is given by 
\beq
\tr\left[\cY_\nu\cY_\nu^T\cY_\nu^*\cY_\nu^\dag\right]\,.
\label{O2Invariant}
\eeq
Summarizing, the basis of invariants for the case of degenerate RH neutrino masses contains all the operators 
listed in Eq.~(\ref{LeptonInvariants}) with $\cA=\unity$, plus the operator in Eq.~(\ref{O2Invariant}).
Moreover, one can show that the degeneracy of the RH neutrino masses leads to a simplification of the general $R$ and $P$ matrices, that can be written as, 
\beq
R=\left(
\begin{array}{cc}
\cosh\omega  & -i\sinh\omega \\
i\sinh\omega  & \cosh\omega \\
\end{array}
\right)\,,\qquad\qquad
P=\left(
\begin{array}{cc}
\cosh2\omega  & -i\sinh2\omega \\
i\sinh2\omega  & \cosh2\omega \\
\end{array}
\right)\,.
\eeq
Consequently, the first condition in Eq.~(\ref{LeptonSolutions}) implies a maximal Majorana phase, 
\beq
\alpha=\pi/4\qquad \text{or}\qquad \alpha=3\pi/4\,,
\label{LeptonSolutionsDegenerate1}
\eeq
for a non-trivial mixing angle. However, this does not imply observability of CP violation at experiments, 
as the relative Majorana phase among the two neutrino eigenvalues is $\pi/2$. Next, the second condition in 
Eq.~(\ref{LeptonSolutions}) can be rewritten as
\beq
\tan2\theta=2\sin2\alpha\dfrac{\sqrt{m_1\,m_2}}{m_1-m_2}\tan2\omega\,,
\label{LeptonSolutionsDegenerate2}
\eeq
which agrees with the results in Ref.~\cite{Alonso:2012fy}, for the particular choice of $\omega$.\footnote{In the notation of Ref.~\cite{Alonso:2012fy}, $\omega$ is defined as $e^\omega\equiv\sqrt{y/y'}$.} 

Eqs.~(\ref{LeptonSolutionsDegenerate1}) and (\ref{LeptonSolutionsDegenerate2}) define a class of extrema of the scalar potential: in particular, a large mixing angle is obtained from almost degenerate masses, while a small angle follows in the hierarchical case. It is, however, 
necessary to discuss the full minimisation of the scalar potential in order to identify the angle configuration 
corresponding to the absolute minimum. In Ref.~\cite{Alonso:2012fy} it was shown that degenerate neutrino masses 
are a good minimum of the scalar potential and therefore one can conclude that the maximal angle solution is 
indeed a good minimum.

Notice that the results in Eqs.~(\ref{LeptonSolutionsDegenerate1}) and (\ref{LeptonSolutionsDegenerate2}) follow 
only considering the last operator in the list of Eq.~(\ref{LeptonInvariants}), with $\cA=\unity$, and in particular 
are not affected by the introduction of the new invariant in Eq.~(\ref{O2Invariant}). Indeed the latter operator 
affects only the neutrino spectrum, having an indirect impact on the results: if the operator 
in Eq.~(\ref{O2Invariant}) is absent, then at the minimum $\omega=0$ and therefore the mixing angle turns out to be vanishing, accordingly to Eq.~(\ref{LeptonSolutionsDegenerate2}) (see Ref.~\cite{Alonso:2012fy} for more details). On the contrary, the operator associated to $\sigma_2$ allows the degenerate mass configuration to be a good minimum 
and therefore, it selects the maximal angle solution as a configuration that minimises the scalar potential.

It is interesting to recover the minima of the scalar potential, using a different parametrisation than the 
Casas-Ibarra one: the bi-unitary parametrisation. The latter consists in decomposing a 
general matrix as a product of a unitary matrix, a diagonal matrix of eigenvalues and a second unitary matrix.  
We will work in the basis in which the RH neutrino and charged lepton mass matrices  are diagonal.
The neutrino Yukawa coupling, vev of the $\cY_\nu$ field, reads in this parametrisation
\beq
Y_\nu\equiv U_L \hat Y_\nu U_R\,,
\label{BiUnitaryNotation}
\eeq
with $U_{L,R}$ being unitary matrices and $\hat Y_\nu={\rm diag}(y_{\nu_1},\,y_{\nu_2})$. The light neutrino mass matrix is then given by
\beq
m_\nu=v^2\,Y_\nu\dfrac{1}{M_N} Y^T_\nu=v^2\, U_L \hat Y_\nu U_R \dfrac{1}{M_N}U^T_R \hat Y_\nu U^T_L\,.
\eeq
Using of the Von Neumann's trace inequality and the freedom of redefining the electron and the muon fields, from the analysis of  $\tr\left[\cY_E\cY_E^\dagger\cY_\nu\cY_\nu^\dagger\right]$, it is straightforward to show that
\beq
U_{L}\propto
\left(
\begin{array}{cc}
 1 & 0 \\
 0 & 1 \\
\end{array}
\right)\,,
\label{LeptonSolutionBiUnitary1}
\eeq
where unphysical phases have been dropped for simplicity. Furthermore, the invariant in Eq.~(\ref{O2Invariant}) leads at the minimum to the following structure for $U_R$
\beq
U_{R}U^T_{R}\propto\left(
\begin{array}{cc}
 0 & 1 \\
 1 & 0 \\
\end{array}
\right)\,,
\label{LeptonSolutionBiUnitary2}
\eeq 
besides the trivial one. The light neutrino mass matrix arising in this context is then given by
\beq
\hat m_\nu=U^T m_\nu U=\dfrac{v^2}{M}\tilde y_{\nu_1} \tilde y_{\nu_2}\,U^T\left(
\begin{array}{cc}
 0 & 1 \\
 1 & 0 \\
\end{array}
\right)U\,,
\label{NeutrinoMassO2}
\eeq
where $\tilde y_{\nu_i}$ contain all the unphysical phases appearing in $U_{L,R}$. As a result, the PMNS matrix reads~\cite{Alonso:2013mca}:
\beq
U=\left(
\begin{array}{cc}
1/\sqrt2  & 1/\sqrt2 \\
-1/\sqrt2  & 1/\sqrt2 \\
\end{array}
\right)
\left(
\begin{array}{cc}
 i &  \\
  & 1 \\
\end{array}
\right)
\eeq
and therefore describes a maximal mixing angle $\theta=\pi/4$ and a maximal relative Majorana phase $2\alpha=\pi/2$, in agreement with the discussion that followed Eq.~(\ref{LeptonSolutionsDegenerate2}). \\

\boldmath
\subsubsection{RH neutrino mass as a flavon: $G^\ell_f=U(2)_{\ell_L}\times U(2)_{E_R}\times U(2)_{N}$.}
\unboldmath

The high predictive power of the degenerate RH neutrino mass case is connected to the symmetry factor $O(2)_{N}$, that allows to write the invariant in Eq.~(\ref{O2Invariant}). One then may ask whether other symmetries produce the same or similar results. In the case of 
\beq
G^\ell_f=U(2)_{\ell_L}\times U(2)_{E_R}\times U(2)_{N}\,
\eeq
the full Lagrangian is invariant only after the promotion of $M_N$ to a dynamical field, properly transforming under $U(2)_{N}$. The operators in Eq.(\ref{LeptonInvariants}) are invariants of $G^\ell_f$, but $\cA=\unity$ is the only possible choice: in particular, the operator in Eq.~(\ref{O2Invariant}) is not an invariant anymore. Moreover, other three additional operators are allowed:
\beq
\tr\left[M_N^*M_N\right]\,,\qquad
\tr\left[\left(M_N^*M_N\right)^2\right]\,, \qquad 
\tr\left[M_N^*M_N\cY_\nu^\dagger\cY_\nu\right]\,,
\eeq
and enter the basis at the renormalisable level.
By using the bi-unitary parametrisation of Eq.~(\ref{BiUnitaryNotation}), it is straightforward to see that 
\beq
\tr\left[M_N^*M_N\cY_\nu^\dagger\cY_\nu\right]\longrightarrow
U_{R}\propto\left(
\begin{array}{cc}
 1 & 0 \\
 0 & 1 \\
\end{array}
\right)\,,
\eeq
or an equivalent configuration. This result, together with Eq.~(\ref{LeptonSolutionBiUnitary1}), leads to a vanishing mixing angle.\\

Summarizing the results for the two-family case, only when the flavour symmetry in the lepton sector includes accounts a $O(2)_{N}$ factor then a maximal angle is a solution at the minimum of the scalar potential, together with a relative Majorana phase of $\pi/2$ and a degenerate light neutrino spectrum.

\subsubsection{Generalisation to the three-family case}

Moving to the realistic scenario of three families of charged leptons and light neutrinos, it is possible to consider either  two or three RH neutrinos (as one of the light neutrino can be massless). In the former case, the interesting case where $G_f^\ell$ accounts for the factor $O(2)_{N}$ is not satisfactory anymore, as the large angle would necessarily  arise in the solar sector (only degenerate masses in tho case) and would lie in the wrong quadrant (see Ref.~\cite{Alonso:2012fy} for further details).

When three RH neutrinos are considered, much of the previous results still hold. If $M_N$ is not promoted to be a dynamical field and neither  degeneracy among the eigenvalues is present, then the symmetry is a trivial generalisation of Eq.~(\ref{GfGenericN}) to three generations and no clear prediction for the mixing angles can be recovered. If $M_N$ is instead a field transforming under an additional $U(3)_{N}$ factor, then the symmetry is that as in Eq.~(\ref{GfGlobal}) and no mixings are predicted at the minimum of the scalar potential. 

All these results are strictly valid considering the scalar potential at the renormalisable level, as the higher order operators are expected to be negligible, under the assumption that the ratio of the flavon vevs and the cutoff of the theory is smaller than 1. Adding non-renormalisable terms to the lepton scalar potential is currently under investigation. 

A second condition has been assumed when extracting the previous results: all the invariants were constructed by means of fields transforming in the bi-fundamental representation of the flavour symmetry. An interesting possibility, that has been already studied for the quark case~\cite{Alonso:2011yg}, is to add fields in the fundamental representation of $G_f^\ell$ and analyse the interplay with the bi-fundamental ones. This naturally happens when two, out of three, RH neutrinos are degenerate in mass~\cite{Alonso:2013mca}: this case corresponds to the flavour symmetry
\beq
G_f^\ell=U(3)_{\ell_L}\times U(3)_{E_R} \times O(2)_{N}\,,
\eeq
and the neutrino Yukawa field $\cY_\nu$ accounts for two components: a doublet and a singlet of $O(2)_{N}$.
Interestingly, two mixing angles can be described in this case: one maximal mixing angle and a maximal Mojarana phase arises in the degenerate sector, as previously, while a second sizable angle is generated due to the interplay with singlet state. This appears as a very promising context as only the third mixing angle remains to be accounted for and it could arise due to small perturbations on the neutrino mass matrix.

\section*{Acknowledgments}
We acknowledge partial support by European Union FP7 ITN INVISIBLES (Marie Curie Actions, PITN-GA-2011-289442), 
CiCYT through the project FPA2009-09017, CAM through the project HEPHACOS P-ESP-00346, European Union FP7 
ITN UNILHC (Marie Curie Actions, PITN-GA-2009-237920), MICINN through the grant BES-2010-037869 and the Juan 
de la Cierva programme (JCI-2011-09244), Italian Ministero dell'Uni\-ver\-si\-t\`a e della Ricerca Scientifica through the COFIN program (PRIN 2008) and the contracts MRTN-CT-2006-035505 and  PITN-GA-2009-237920 (UNILHC). Finally we thank the organizers of the Moriond EW conference for the kind invitation and for their efforts in organizing this enjoyable meeting.

\section*{References}
\providecommand{\href}[2]{#2}\begingroup\raggedright\endgroup


\begin{thebibliography}{10}

\bibitem{Aad:2012tfa}
{\bf ATLAS Collaboration} Collaboration, G.~Aad {\em et.~al.},  Phys.Lett. {\bf
  B716} (2012) 1--29, [\href{http://xxx.lanl.gov/abs/1207.7214}{{\tt
  arXiv:1207.7214}}].

\bibitem{Chatrchyan:2012ufa}
{\bf CMS Collaboration} Collaboration, S.~Chatrchyan {\em et.~al.},  Phys.Lett.
  {\bf B716} (2012) 30--61, [\href{http://xxx.lanl.gov/abs/1207.7235}{{\tt
  arXiv:1207.7235}}].

\bibitem{Giudice:2007fh}
G.~Giudice, C.~Grojean, A.~Pomarol, and R.~Rattazzi,  JHEP {\bf 0706} (2007)
  045, [\href{http://xxx.lanl.gov/abs/hep-ph/0703164}{{\tt hep-ph/0703164}}].

\bibitem{Grinstein:2007iv}
B.~Grinstein and M.~Trott,  Phys.Rev. {\bf D76} (2007) 073002,
  [\href{http://xxx.lanl.gov/abs/0704.1505}{{\tt arXiv:0704.1505}}].

\bibitem{Contino:2010mh}
R.~Contino, C.~Grojean, M.~Moretti, F.~Piccinini, and R.~Rattazzi,  JHEP {\bf
  1005} (2010) 089, [\href{http://xxx.lanl.gov/abs/1002.1011}{{\tt
  arXiv:1002.1011}}].

\bibitem{Alonso:2012jc}
R.~Alonso, M.~Gavela, L.~Merlo, S.~Rigolin, and J.~Yepes,  JHEP {\bf 1206}
  (2012) 076, [\href{http://xxx.lanl.gov/abs/1201.1511}{{\tt
  arXiv:1201.1511}}].

\bibitem{Azatov:2012bz}
A.~Azatov, R.~Contino, and J.~Galloway,  JHEP {\bf 1204} (2012) 127,
  [\href{http://xxx.lanl.gov/abs/1202.3415}{{\tt arXiv:1202.3415}}].

\bibitem{Corbett:2012dm}
T.~Corbett, O.~Eboli, J.~Gonzalez-Fraile, and M.~Gonzalez-Garcia,  Phys.Rev.
  {\bf D86} (2012) 075013, [\href{http://xxx.lanl.gov/abs/1207.1344}{{\tt
  arXiv:1207.1344}}].

\bibitem{Espinosa:2012im}
J.~Espinosa, C.~Grojean, M.~Muhlleitner, and M.~Trott,
  \href{http://xxx.lanl.gov/abs/1207.1717}{{\tt arXiv:1207.1717}}.

\bibitem{Corbett:2012ja}
T.~Corbett, O.~Eboli, J.~Gonzalez-Fraile, and M.~Gonzalez-Garcia,  Phys.Rev.
  {\bf D87} (2013) 015022, [\href{http://xxx.lanl.gov/abs/1211.4580}{{\tt
  arXiv:1211.4580}}].

\bibitem{Alonso:2012px}
R.~Alonso, M.~Gavela, L.~Merlo, S.~Rigolin, and J.~Yepes,  Phys.Lett. {\bf
  B722} (2013) 330--335, [\href{http://xxx.lanl.gov/abs/1212.3305}{{\tt
  arXiv:1212.3305}}].

\bibitem{Alonso:2012pz}
R.~Alonso, M.~Gavela, L.~Merlo, S.~Rigolin, and J.~Yepes,  Phys.Rev. {\bf D87}
  (2013) 055019, [\href{http://xxx.lanl.gov/abs/1212.3307}{{\tt
  arXiv:1212.3307}}].

\bibitem{Azatov}
A.~Azatov \!\!, Talk given at the XLVIIIth Rencontres de Moriond session
  devoted to ELEC- TROWEAK INTERACTIONS AND UNIFIED THEORIES, La Thuile
  (Italy), 2-9 March 2013.

\bibitem{Corbett:2013hia}
T.~Corbett, O.~Eboli, J.~Gonzalez-Fraile, and M.~Gonzalez-Garcia,
  \href{http://xxx.lanl.gov/abs/1306.0006}{{\tt arXiv:1306.0006}}.

\bibitem{Alonso:2013voa}
R.~Alonso, M.~Gavela, L.~Merlo, S.~Rigolin, and J.~Yepes,
  \href{http://xxx.lanl.gov/abs/1304.5937}{{\tt arXiv:1304.5937}}.

\bibitem{Ciuchini:2000de}
M.~Ciuchini, G.~D'Agostini, E.~Franco, V.~Lubicz, G.~Martinelli, {\em et.~al.},
   JHEP {\bf 0107} (2001) 013,
  [\href{http://xxx.lanl.gov/abs/hep-ph/0012308}{{\tt hep-ph/0012308}}].

\bibitem{Charles:2004jd}
{\bf CKMfitter Group} Collaboration, J.~Charles {\em et.~al.},  Eur.Phys.J.
  {\bf C41} (2005) 1--131, [\href{http://xxx.lanl.gov/abs/hep-ph/0406184}{{\tt
  hep-ph/0406184}}].

\bibitem{Alonso:2012ji}
R.~Alonso, M.~Dhen, M.~Gavela, and T.~Hambye,  JHEP {\bf 1301} (2013) 118,
  [\href{http://xxx.lanl.gov/abs/1209.2679}{{\tt arXiv:1209.2679}}].

\bibitem{Hungerford:2009zz}
{\bf COMET Collaboration} Collaboration, E.~V. Hungerford,  AIP Conf.Proc. {\bf
  1182} (2009) 694--697.

\bibitem{Cui:2009zz}
{\bf COMET Collaboration} Collaboration, Y.~Cui {\em et.~al.}, .

\bibitem{Carey:2008zz}
{\bf Mu2e Collaboration} Collaboration, R.~Carey {\em et.~al.}, .

\bibitem{Kutschke:2011ux}
R.~K. Kutschke,  \href{http://xxx.lanl.gov/abs/1112.0242}{{\tt
  arXiv:1112.0242}}.

\bibitem{Kurup:2011zza}
{\bf COMET Collaboration} Collaboration, A.~Kurup,  Nucl.Phys.Proc.Suppl. {\bf
  218} (2011) 38--43.

\bibitem{Deppisch:2010fr}
F.~F. Deppisch and A.~Pilaftsis,  Phys.Rev. {\bf D83} (2011) 076007,
  [\href{http://xxx.lanl.gov/abs/1012.1834}{{\tt arXiv:1012.1834}}].

\bibitem{Ilakovac:2009jf}
A.~Ilakovac and A.~Pilaftsis,  Phys.Rev. {\bf D80} (2009) 091902,
  [\href{http://xxx.lanl.gov/abs/0904.2381}{{\tt arXiv:0904.2381}}].

\bibitem{Froggatt:1978nt}
C.~Froggatt and H.~B. Nielsen,  Nucl.Phys. {\bf B147} (1979) 277.

\bibitem{Buchmuller:2011tm}
W.~Buchmuller, V.~Domcke, and K.~Schmitz,  JHEP {\bf 1203} (2012) 008,
  [\href{http://xxx.lanl.gov/abs/1111.3872}{{\tt arXiv:1111.3872}}].

\bibitem{Altarelli:2012ia}
G.~Altarelli, F.~Feruglio, I.~Masina, and L.~Merlo,  JHEP {\bf 1211} (2012)
  139, [\href{http://xxx.lanl.gov/abs/1207.0587}{{\tt arXiv:1207.0587}}].

\bibitem{Calibbi:2012yj}
L.~Calibbi, Z.~Lalak, S.~Pokorski, and R.~Ziegler,  JHEP {\bf 1206} (2012) 018,
  [\href{http://xxx.lanl.gov/abs/1203.1489}{{\tt arXiv:1203.1489}}].

\bibitem{Calibbi:2012at}
L.~Calibbi, Z.~Lalak, S.~Pokorski, and R.~Ziegler,
  \href{http://xxx.lanl.gov/abs/1204.1275}{{\tt arXiv:1204.1275}}.

\bibitem{Altarelli:2010gt}
G.~Altarelli and F.~Feruglio,  Rev. Mod. Phys. {\bf 82} (2010) 2701--2729,
  [\href{http://xxx.lanl.gov/abs/1002.0211}{{\tt arXiv:1002.0211}}].

\bibitem{Ishimori:2010au}
H.~Ishimori {\em et.~al.},  Prog. Theor. Phys. Suppl. {\bf 183} (2010) 1--163,
  [\href{http://xxx.lanl.gov/abs/1003.3552}{{\tt arXiv:1003.3552}}].

\bibitem{Grimus:2011fk}
W.~Grimus and P.~O. Ludl,  \href{http://xxx.lanl.gov/abs/1110.6376}{{\tt
  arXiv:1110.6376}}.

\bibitem{Altarelli:2012ss}
G.~Altarelli, F.~Feruglio, and L.~Merlo,  Fortsch.Phys. {\bf 61} (2013)
  507--534, [\href{http://xxx.lanl.gov/abs/1205.5133}{{\tt arXiv:1205.5133}}].

\bibitem{Bazzocchi:2012st}
F.~Bazzocchi and L.~Merlo,  Fortsch.Phys. {\bf 61} (2013) 571--596,
  [\href{http://xxx.lanl.gov/abs/1205.5135}{{\tt arXiv:1205.5135}}].

\bibitem{Feruglio:2007uu}
F.~Feruglio, C.~Hagedorn, Y.~Lin, and L.~Merlo,  Nucl. Phys. {\bf B775} (2007)
  120--142, [\href{http://xxx.lanl.gov/abs/hep-ph/0702194}{{\tt
  hep-ph/0702194}}].

\bibitem{Feruglio:2008ht}
F.~Feruglio, C.~Hagedorn, Y.~Lin, and L.~Merlo,  Nucl.Phys. {\bf B809} (2009)
  218--243, [\href{http://xxx.lanl.gov/abs/0807.3160}{{\tt arXiv:0807.3160}}].

\bibitem{Ishimori:2008au}
H.~Ishimori, T.~Kobayashi, Y.~Omura, and M.~Tanimoto,  JHEP {\bf 0812} (2008)
  082, [\href{http://xxx.lanl.gov/abs/0807.4625}{{\tt arXiv:0807.4625}}].

\bibitem{Ishimori:2009ew}
H.~Ishimori, T.~Kobayashi, H.~Okada, Y.~Shimizu, and M.~Tanimoto,  JHEP {\bf
  0912} (2009) 054, [\href{http://xxx.lanl.gov/abs/0907.2006}{{\tt
  arXiv:0907.2006}}].

\bibitem{Feruglio:2009iu}
F.~Feruglio, C.~Hagedorn, and L.~Merlo,  JHEP {\bf 1003} (2010) 084,
  [\href{http://xxx.lanl.gov/abs/0910.4058}{{\tt arXiv:0910.4058}}].

\bibitem{Feruglio:2009hu}
F.~Feruglio, C.~Hagedorn, Y.~Lin, and L.~Merlo,  Nucl.Phys. {\bf B832} (2010)
  251--288, [\href{http://xxx.lanl.gov/abs/0911.3874}{{\tt arXiv:0911.3874}}].

\bibitem{Toorop:2010ex}
R.~de~Adelhart~Toorop, F.~Bazzocchi, L.~Merlo, and A.~Paris,  JHEP {\bf 1103}
  (2011) 035, [\href{http://xxx.lanl.gov/abs/1012.1791}{{\tt
  arXiv:1012.1791}}].

\bibitem{Toorop:2010kt}
R.~de~Adelhart~Toorop, F.~Bazzocchi, L.~Merlo, and A.~Paris,  JHEP {\bf 1103}
  (2011) 040, [\href{http://xxx.lanl.gov/abs/1012.2091}{{\tt
  arXiv:1012.2091}}].

\bibitem{Merlo:2011hw}
L.~Merlo, S.~Rigolin, and B.~Zaldivar,  JHEP {\bf 11} (2011) 047,
  [\href{http://xxx.lanl.gov/abs/1108.1795}{{\tt arXiv:1108.1795}}].

\bibitem{Altarelli:2012bn}
G.~Altarelli, F.~Feruglio, L.~Merlo, and E.~Stamou,  JHEP {\bf 1208} (2012)
  021, [\href{http://xxx.lanl.gov/abs/1205.4670}{{\tt arXiv:1205.4670}}].

\bibitem{Altarelli:2009gn}
G.~Altarelli, F.~Feruglio, and L.~Merlo,  JHEP {\bf 0905} (2009) 020,
  [\href{http://xxx.lanl.gov/abs/0903.1940}{{\tt arXiv:0903.1940}}].

\bibitem{Toorop:2010yh}
R.~de~Adelhart~Toorop, F.~Bazzocchi, and L.~Merlo,  JHEP {\bf 1008} (2010) 001,
  [\href{http://xxx.lanl.gov/abs/1003.4502}{{\tt arXiv:1003.4502}}].

\bibitem{Varzielas:2010mp}
I.~de~Medeiros~Varzielas and L.~Merlo,  JHEP {\bf 02} (2011) 062,
  [\href{http://xxx.lanl.gov/abs/1011.6662}{{\tt arXiv:1011.6662}}].

\bibitem{Ge:2011ih}
S.-F. Ge, D.~A. Dicus, and W.~W. Repko,  Phys.Lett. {\bf B702} (2011) 220--223,
  [\href{http://xxx.lanl.gov/abs/1104.0602}{{\tt arXiv:1104.0602}}].

\bibitem{Ma:2011yi}
E.~Ma and D.~Wegman,  Phys. Rev. Lett. {\bf 107} (2011) 061803,
  [\href{http://xxx.lanl.gov/abs/1106.4269}{{\tt arXiv:1106.4269}}].

\bibitem{Meloni:2011fx}
D.~Meloni,  JHEP {\bf 10} (2011) 010,
  [\href{http://xxx.lanl.gov/abs/1107.0221}{{\tt arXiv:1107.0221}}].

\bibitem{Morisi:2011pm}
S.~Morisi, K.~M. Patel, and E.~Peinado,  Phys. Rev. {\bf D84} (2011) 053002,
  [\href{http://xxx.lanl.gov/abs/1107.0696}{{\tt arXiv:1107.0696}}].

\bibitem{Toorop:2011jn}
R.~d.~A. Toorop, F.~Feruglio, and C.~Hagedorn,  Phys.Lett. {\bf B703} (2011)
  447--451, [\href{http://xxx.lanl.gov/abs/1107.3486}{{\tt arXiv:1107.3486}}].

\bibitem{King:2011zj}
S.~F. King and C.~Luhn,  JHEP {\bf 09} (2011) 042,
  [\href{http://xxx.lanl.gov/abs/1107.5332}{{\tt arXiv:1107.5332}}].

\bibitem{Hernandez:2012ra}
D.~Hernandez and A.~Y. Smirnov,  \href{http://xxx.lanl.gov/abs/1204.0445}{{\tt
  arXiv:1204.0445}}.

\bibitem{Feruglio:2012cw}
F.~Feruglio, C.~Hagedorn, and R.~Ziegler,
  \href{http://xxx.lanl.gov/abs/1211.5560}{{\tt arXiv:1211.5560}}.

\bibitem{Holthausen:2012dk}
M.~Holthausen, M.~Lindner, and M.~A. Schmidt,
  \href{http://xxx.lanl.gov/abs/1211.6953}{{\tt arXiv:1211.6953}}.

\bibitem{Hernandez:2012sk}
D.~Hernandez and A.~Y. Smirnov,  \href{http://xxx.lanl.gov/abs/1212.2149}{{\tt
  arXiv:1212.2149}}.

\bibitem{Feruglio:2013hia}
F.~Feruglio, C.~Hagedorn, and R.~Ziegler,
  \href{http://xxx.lanl.gov/abs/1303.7178}{{\tt arXiv:1303.7178}}.

\bibitem{Chivukula:1987py}
R.~S. Chivukula and H.~Georgi,  Phys.Lett. {\bf B188} (1987) 99.

\bibitem{Hall:1990ac}
L.~J. Hall and L.~Randall,  Phys. Rev. Lett. {\bf 65} (1990) 2939--2942.

\bibitem{D'Ambrosio:2002ex}
G.~D'Ambrosio, G.~Giudice, G.~Isidori, and A.~Strumia,  Nucl.Phys. {\bf B645}
  (2002) 155--187, [\href{http://xxx.lanl.gov/abs/hep-ph/0207036}{{\tt
  hep-ph/0207036}}].

\bibitem{Cirigliano:2005ck}
V.~Cirigliano, B.~Grinstein, G.~Isidori, and M.~B. Wise,  Nucl.Phys. {\bf B728}
  (2005) 121--134, [\href{http://xxx.lanl.gov/abs/hep-ph/0507001}{{\tt
  hep-ph/0507001}}].

\bibitem{Davidson:2006bd}
S.~Davidson and F.~Palorini,  Phys.Lett. {\bf B642} (2006) 72--80,
  [\href{http://xxx.lanl.gov/abs/hep-ph/0607329}{{\tt hep-ph/0607329}}].

\bibitem{Kagan:2009bn}
A.~L. Kagan, G.~Perez, T.~Volansky, and J.~Zupan,  Phys. Rev. {\bf D80} (2009)
  076002, [\href{http://xxx.lanl.gov/abs/0903.1794}{{\tt arXiv:0903.1794}}].

\bibitem{Gavela:2009cd}
M.~Gavela, T.~Hambye, D.~Hernandez, and P.~Hernandez,  JHEP {\bf 0909} (2009)
  038, [\href{http://xxx.lanl.gov/abs/0906.1461}{{\tt arXiv:0906.1461}}].

\bibitem{Feldmann:2009dc}
T.~Feldmann, M.~Jung, and T.~Mannel,  Phys.Rev. {\bf D80} (2009) 033003,
  [\href{http://xxx.lanl.gov/abs/0906.1523}{{\tt arXiv:0906.1523}}].

\bibitem{Alonso:2011yg}
R.~Alonso, M.~Gavela, L.~Merlo, and S.~Rigolin,  JHEP {\bf 1107} (2011) 012,
  [\href{http://xxx.lanl.gov/abs/1103.2915}{{\tt arXiv:1103.2915}}].

\bibitem{Alonso:2011jd}
R.~Alonso, G.~Isidori, L.~Merlo, L.~A. Munoz, and E.~Nardi,  JHEP {\bf 1106}
  (2011) 037, [\href{http://xxx.lanl.gov/abs/1103.5461}{{\tt
  arXiv:1103.5461}}].

\bibitem{Alonso:2012fy}
R.~Alonso, M.~Gavela, D.~Hernandez, and L.~Merlo,  Phys.Lett. {\bf B715} (2012)
  194--198, [\href{http://xxx.lanl.gov/abs/1206.3167}{{\tt arXiv:1206.3167}}].

\bibitem{Lalak:2010bk}
Z.~Lalak, S.~Pokorski, and G.~G. Ross,  JHEP {\bf 1008} (2010) 129,
  [\href{http://xxx.lanl.gov/abs/1006.2375}{{\tt arXiv:1006.2375}}].

\bibitem{Fitzpatrick:2007sa}
A.~L. Fitzpatrick, G.~Perez, and L.~Randall,  Phys.Rev.Lett. {\bf 100} (2008)
  171604, [\href{http://xxx.lanl.gov/abs/0710.1869}{{\tt arXiv:0710.1869}}].

\bibitem{Grinstein:2010ve}
B.~Grinstein, M.~Redi, and G.~Villadoro,  JHEP {\bf 1011} (2010) 067,
  [\href{http://xxx.lanl.gov/abs/1009.2049}{{\tt arXiv:1009.2049}}].

\bibitem{Buras:2011wi}
A.~J. Buras, M.~V. Carlucci, L.~Merlo, and E.~Stamou,  JHEP {\bf 1203} (2012)
  088, [\href{http://xxx.lanl.gov/abs/1112.4477}{{\tt arXiv:1112.4477}}].

\bibitem{Barbieri:2011ci}
R.~Barbieri, G.~Isidori, J.~Jones-Perez, P.~Lodone, and D.~M. Straub,  Eur.
  Phys. J. {\bf C71} (2011) 1725,
  [\href{http://xxx.lanl.gov/abs/1105.2296}{{\tt arXiv:1105.2296}}].

\bibitem{Blankenburg:2012nx}
G.~Blankenburg, G.~Isidori, and J.~Jones-Perez,  Eur.Phys.J. {\bf C72} (2012)
  2126, [\href{http://xxx.lanl.gov/abs/1204.0688}{{\tt arXiv:1204.0688}}].

\bibitem{Lopez-Honorez:2013wla}
L.~Lopez-Honorez and L.~Merlo,  Phys.Lett. {\bf B722} (2013) 135--143,
  [\href{http://xxx.lanl.gov/abs/1303.1087}{{\tt arXiv:1303.1087}}].

\bibitem{Isidori:2010kg}
G.~Isidori, Y.~Nir, and G.~Perez,  Ann. Rev. Nucl. Part. Sci. {\bf 60} (2010)
  355, [\href{http://xxx.lanl.gov/abs/1002.0900}{{\tt arXiv:1002.0900}}].

\bibitem{Anselm:1996jm}
A.~Anselm and Z.~Berezhiani,  Nucl.Phys. {\bf B484} (1997) 97--123,
  [\href{http://xxx.lanl.gov/abs/hep-ph/9605400}{{\tt hep-ph/9605400}}].

\bibitem{Barbieri:1999km}
R.~Barbieri, L.~J. Hall, G.~L. Kane, and G.~G. Ross,
  \href{http://xxx.lanl.gov/abs/hep-ph/9901228}{{\tt hep-ph/9901228}}.

\bibitem{Berezhiani:2001mh}
Z.~Berezhiani and A.~Rossi,  Nucl.Phys.Proc.Suppl. {\bf 101} (2001) 410--420,
  [\href{http://xxx.lanl.gov/abs/hep-ph/0107054}{{\tt hep-ph/0107054}}].

\bibitem{Harrison:2005dj}
P.~Harrison and W.~Scott,  Phys.Lett. {\bf B628} (2005) 93,
  [\href{http://xxx.lanl.gov/abs/hep-ph/0508012}{{\tt hep-ph/0508012}}].

\bibitem{Alonso:2013mca}
R.~Alonso, M.~Gavela, D.~Hernandez, L.~Merlo, and S.~Rigolin,  JHEP {\bf 1308}
  (2013) 069, [\href{http://xxx.lanl.gov/abs/1306.5922}{{\tt
  arXiv:1306.5922}}].

\bibitem{Feldmann:2010yp}
T.~Feldmann,  JHEP {\bf 04} (2011) 043,
  [\href{http://xxx.lanl.gov/abs/1010.2116}{{\tt arXiv:1010.2116}}].

\bibitem{Guadagnoli:2011id}
D.~Guadagnoli, R.~N. Mohapatra, and I.~Sung,  JHEP {\bf 04} (2011) 093,
  [\href{http://xxx.lanl.gov/abs/1103.4170}{{\tt arXiv:1103.4170}}].

\bibitem{Buras:2011zb}
A.~J. Buras, L.~Merlo, and E.~Stamou,  JHEP {\bf 1108} (2011) 124,
  [\href{http://xxx.lanl.gov/abs/1105.5146}{{\tt arXiv:1105.5146}}].

\bibitem{Nardi:2011st}
E.~Nardi,  Phys.Rev. {\bf D84} (2011) 036008,
  [\href{http://xxx.lanl.gov/abs/1105.1770}{{\tt arXiv:1105.1770}}].

\bibitem{Espinosa:2012uu}
J.~R. Espinosa, C.~S. Fong, and E.~Nardi,  JHEP {\bf 1302} (2013) 137,
  [\href{http://xxx.lanl.gov/abs/1211.6428}{{\tt arXiv:1211.6428}}].

\bibitem{Casas:2001sr}
J.~Casas and A.~Ibarra,  Nucl.Phys. {\bf B618} (2001) 171--204,
  [\href{http://xxx.lanl.gov/abs/hep-ph/0103065}{{\tt hep-ph/0103065}}].

\end{thebibliography}
\end{document}